\documentstyle[12pt,epsfig]{article}

\renewcommand{\a}{\alpha}

\newcommand{\g}{\gamma}
\newcommand{\la}{\lambda}
\newcommand{\G}{\Gamma}
\renewcommand{\S}{\Sigma}
\newcommand{\La}{\Lambda}
\newcommand{\eps}{\epsilon}

\newcommand{\bea}{\begin{eqnarray}}
\newcommand{\eea}{\end{eqnarray}}
\newcommand{\beq}{\begin{equation}}
\newcommand{\eeq}{\end{equation}}
\newcommand{\nn}{\nonumber}
\newcommand{\fr}{\frac}
\newcommand{\hl}{\hline}

\newcommand{\ra}{\rightarrow}
\begin{document}
\begin{titlepage}


\begin{flushright} {FTUV/99-75 \\ IFIC/99-78}
\end{flushright}

\vskip 2cm

\centerline{\LARGE \bf
Electromagnetic Decays of Heavy Baryons
}
\vskip 1cm
\centerline{M.C. Ba\~nuls, A. Pich, I. Scimemi}
\vskip 0.5cm
\centerline{Departament de F\'{\i}sica Te\`orica, IFIC, 
  Universitat de Val\`encia -- CSIC}
\centerline{ E-46100 Burjassot (Val\`encia), Spain}

\begin{abstract}
The electromagnetic decays of the ground state baryon multiplets with one
heavy quark are calculated using  Heavy Hadron Chiral Perturbation Theory.
The M1 and E2 amplitudes for 
$S^{*}\ra S \g $, $S^{*} \ra T \g$ and $S \ra T \g$
are separately computed.
All M1 transitions are calculated up to  ${\cal O}(1/\Lambda_\chi^2)$.
The E2 amplitudes contribute at the same order for $S^{*}\ra S \g $, while
for $S^{*} \ra T \g$ they first appear at  
${\cal O}(1/(m_Q \Lambda_\chi^2))$
and for $S \ra T \g$ are completely negligible.
The renormalization  of the chiral loops is discussed  and 
relations among different decay amplitudes are derived. 
We find that  
chiral loops involving electromagnetic interactions of the light
pseudoscalar mesons
 provide a sizable enhancement of these decay widths.
Furthermore, we obtain an absolute prediction for
$\G(\Xi^{0'(*)}_c\ra \Xi^{0}_c \g) $ and
$\G(\Xi^{-'(*)}_b\ra \Xi^{-}_b \g) $. 
Our results are compared to other estimates existing in the literature.  
\end{abstract}

\end{titlepage}

\newpage

\section{Introduction}

In some kinematical regions, which are not far from the chiral 
and heavy quark limits, both Chiral Perturbation~\cite{pich}
and Heavy Quark Effective  Theories (HQET)~\cite{pichLH} 
can be simultaneously used.
In the $m_{Q}\rightarrow \infty$ limit, baryons containing a heavy quark, 
can emit and absorb light pseudoscalar mesons without changing its velocity $v$. 
In Heavy Hadron Chiral Perturbation Theory (HHCPT) one constructs 
an effective Lagrangian whose basic fields are heavy hadrons and 
light mesons~\cite{wise}-\cite{cho}. In ref.~\cite{chogeo}, 
the formalism is extended to include also electromagnetism.

We use this hybrid effective Lagrangian to calculate 
the electromagnetic decay width 
of the ground state baryons containing a $c$ or a $b$ quark.
We consider the decays $S^{*}\ra S \g $ and $S^{(*)} \ra T \g$.
For most of these decays the available phase space is small,
so that  the  emission of a pion is suppressed or even forbidden 
and the electromagnetic process becomes relevant.  
Some of these decays are starting to be measured~\cite{expcleo},
which makes necessary to perform a detailed theoretical analysis.

Some theoretical calculations of these decays can be already found
in the literature.
The  ${\cal O}(1/\La_\chi)$  amplitudes
were first computed in ref.~\cite{chogeo}, using HHCPT.
A more detailed analysis was presented in ref.~\cite{chengnu},
where the widths  $\G (S_c\ra T_c\g)$  are estimated using
heavy--quark and chiral symmetries
implemented within the non-relativistic quark model.
A similar procedure is followed in ref.~\cite{chori}, where 
the heavy--quark symmetry is supplemented with light--diquark  symmetries
to calculate the widths
$\G(\S^{+}_c\ra \La^{+}_c \g)$ and $\G(\S^{*}_{c,b}\ra \S_{c,b} \g)$.
The authors of ref.~\cite{ivan} apply the relativistic quark model
to predict the electromagnetic decays  
$\G (S^{(*)}_c\ra T_c\g)$ and $\G(\S^{+*}_c\ra \S^{+}_c \g)$.
In ref.~\cite{redai},  $\G(\S^{*}_b\ra \S_b \g)$ and
$\G(\S^{0(*)}_b\ra \La^{0}_b \g)$ are computed with 
light cone QCD sum rules at leading order in HQET.
All these references consider only transitions of the  M1 type.
Finally, ref.~\cite{savanu} estimates the ratio of the  E2 and M1 amplitudes
for  $\G(\S^{+*}_c\ra \La^{+}_c \g)$.

Here, we  study all possible $S^{(*)} \ra T \g$ and 
$S^{*} \ra S \g$ decays in the context of HHCPT, considering both
M1 and E2 transitions.
Section~\ref{sec:form} collects the
HHCPT formalism as introduced in ref.~\cite{chogeo}:
the effective fields representing $S$ and $T$ baryons, 
the lowest order chiral Lagrangian and the ${\cal O}(1/m_Q)$ and
${\cal O}(1/\La_\chi)$ terms.
In order to renormalize the resulting
chiral loops, the introduction of higher--order operators 
with unknown couplings is required.
In the case of $S^* \ra S \g$, 
we calculate all contributions  up  to   ${\cal O}(1/\La_\chi^2)$ for 
M1 and E2 transitions.
We find that all divergences and scale dependence  
can be absorbed in the redefinition of one  ${\cal O}(1/\Lambda_\chi)$
coupling for each type of process (M1, E2).
These results are presented in section~\ref{sec:S}.
Section~\ref{sec:T} describes the analogous calculation for $S^* \ra T \g$;
in this case, the E2 contribution has to be computed 
up to  ${\cal O}(1/m_Q \La_\chi^2)$, which requires
two additional couplings.
The decays $S\ra T \g$ are analyzed in section~\ref{sec:st};
as in the previous cases the M1 amplitude is calculated up to
 ${\cal O}(1/\La_\chi^2)$, while the E2 contribution
is found to be ${\cal O}(1/m_Q^3 \La_\chi^2)$ and thus extremely suppressed.
In each section we derive relations among amplitudes for different baryons 
within the same multiplet and between charm and bottom baryons.
These relations are valid at lowest order in HHCPT and we prove that they
still hold after one--loop chiral corrections are included.
Comparing our expectation for the widths to the leading order HHCPT estimate,
we find that the infrared effect due to
 electromagnetic interactions of light pseudoscalar mesons
 can greatly enhance these widths.
This is particularly true for  the E2 contributions which are found to be
infrared divergent in the exact chiral limit.
We also  give some comments on   results  existing in the literature.
Finally, section~\ref{sec:fin} summarizes our conclusions.

\section{HHCPT formalism for magnetic moments}
\label{sec:form}

The light degrees of freedom in the ground  state  of a 
baryon  with one heavy quark can be either in
a $s_l=0$ or in a $s_l=1$ configuration.
The first one corresponds  to $J^P=\frac{1}{2}^+$ baryons, which 
are annihilated by $T_i(v)$ fields transforming
as a $\bar{\bf3}$ under the chiral subgroup $SU(3)_{L+R}$ and as a doublet under 
the HQET $SU(2)_{v}$.
In the second case, $s_l=1$, the spin of the heavy quark and the light
degrees of freedom combine together to form
$J^P=\frac{3}{2}^+$ and $J=\frac{1}{2}^+$ baryons, which are degenerate
in  mass  in the $m_Q\rightarrow \infty$ limit.
The spin--$\frac{3}{2}$ ones are annihilated
by the Rarita--Schwinger field $S_{\mu}^{* ij}(v)$,
 while the spin--$\frac{1}{2}$
baryons are destroyed by the Dirac field $S^{ ij}(v)$. 
It is very useful to combine
both operators into the so-called superfield \cite{georgi,falk} 
\bea
S_{\mu}^{ij} (v)&=& \sqrt{\frac{1}{3}} \, (\gamma_{\mu} + v_{\mu}) 
  \,\gamma^5\,  S^{ij} (v) +  S_{\mu}^{* ij}(v) \ ,
\nonumber \\
\bar{S}_{ij}^{\mu} (v) &=& - \sqrt{\frac{1}{3}} \,\bar{S}_{ij}(v) 
  \,\gamma^5\,  (\gamma^{\mu} + v^{\mu})+ \bar{S}_{ij}^{* \mu}(v) \ ,
\eea
which transforms as a {\bf 6} under $SU(3)_{L+R}$ and as a doublet 
under $SU(2)_{v}$ and is symmetric in the $i$, $j$ indices.

The particle assignment for the  $J= 1/2$ charmed baryons of 
the $\bar{\bf 3}$ and {\bf 6} representations is
\beq
(T_1,T_2,T_3)\, =\, (\Xi^0_c,-\Xi^+_c,\Lambda^+_c) \ ,
\eeq
\beq
S^{i j} \, =\, \left ( 
\begin{array}{ccc} 
\Sigma^{++}_c& \sqrt{\frac{1}{2}} \Sigma^+_c & \sqrt{\frac{1}{2}} \Xi^{+'}_c \\
\sqrt{\frac{1}{2}} \Sigma^{+}_c &\Sigma^0_c & \sqrt{\frac{1}{2}} \Xi^{0'}_c \\
\sqrt{\frac{1}{2}} \Xi^{+'}_c & \sqrt{\frac{1}{2}} \Xi^{0'}_c & \Omega^0_c 
\end{array}
\right )\ ,
\label{eq:sba}
\eeq
and the corresponding bottom baryons are
\beq
(T_1,T_2,T_3) \, =\, (\Xi^-_b,-\Xi^0_b,\Lambda^0_b) \ ,
\eeq
\beq
S^{i j} \, = \,\left( 
\begin{array}{ccc} 
\Sigma^{+}_b& \sqrt{\frac{1}{2}} \Sigma^0_b & \sqrt{\frac{1}{2}} \Xi^{0'}_b \\
\sqrt{\frac{1}{2}} \Sigma^{0}_b &\Sigma^-_b & \sqrt{\frac{1}{2}} \Xi^{-'}_b \\
\sqrt{\frac{1}{2}} \Xi^{0'}_b & \sqrt{\frac{1}{2}} \Xi^{-'}_b & \Omega^-_b 
\end{array}
\right ) \ .
\label{eq:sbab}
\eeq
The  $J=3/2$  partners  of the   baryons in Eq.~(\ref{eq:sba}) and 
Eq.~(\ref{eq:sbab}) have the same $SU(3)_V$ assignments in $S_\mu^{*ij}$.

Goldstone bosons are parametrized as
\beq
{\Phi} \, =\, \left( 
\begin{array}{ccc}
\sqrt{\frac{1}{2}} \pi^0 + \sqrt{\frac{1}{6}} \eta & \pi^+ & K^+ \\
\pi^- & - \sqrt{\frac{1}{2}} \pi^0 +\sqrt{\frac{1}{6}} \eta & K^0 \\
K^- & \bar{K}^0 & -\sqrt{\frac{2}{3}} \eta 
\end{array}
\right)  \ , \label{pi}
\eeq
and appear in the Lagrangian via the exponential representation
$\xi \equiv\exp( i \Phi /\sqrt{2} f_{\pi})$,  being $f_\pi\sim 93$ MeV the 
pion decay constant.

The lowest--order chiral Lagrangian describing the soft hadronic and 
electromagnetic interactions of these baryons in the infinite 
heavy quark mass limit is given by \cite{chogeo}
\bea
{\cal L}^{(0)} &=& -i \,\bar{S}_{ij}^{\mu}\, (v \cdot D) S_{\mu}^{ij} +
\Delta_{ST} \,\bar{S}_{ij}^{\mu}\, S_{\mu}^{ij} + 
i\, \bar{T}^i\, (v \cdot D) T_i 
\nonumber \\ &&\mbox{}
+ i \, g_2 \,\varepsilon_{\mu \nu \sigma \lambda} \,
\bar{S}_{ik}^{\mu} v^{\nu} (\xi^{\sigma})_j^i (S^{\lambda})^{jk} 
+ g_3 \,\left[\epsilon_{ijk} \,\bar{T}^i (\xi^{\mu})_l^j S_{\mu}^{kl}
+ \epsilon^{ijk}\, \bar{S}_{kl}^{\mu} (\xi_{\mu})_j^l T_i\right]\ .
\label{eq:lagos}
\eea
In this formula, the heavy--baryon covariant derivatives are
\bea
D^{\mu} S_{\nu}^{ij}&=&\partial^{\mu}S_{\nu}^{ij}+(\G^{\mu})_k^i S_{\nu}^{kj}+
(\G^{\mu})_k^j S_{\nu}^{ik}-i e {\cal A}^{\mu}\,
[Q_Q S_{\nu}^{ij}+Q_k^i S_{\nu}^{kj}+Q_k^j S_{\nu}^{ik}] \ , 
\nonumber \\
D^{\mu} T_i &=& \partial^{\mu} T_i - T_j (\G^{\mu})_i^j 
- i e {\cal A}^{\mu}\, [Q_Q T_i -T_j Q_i^j] \ ,
\eea
where ${\cal A}^{\mu}$ is the electromagnetic field,
$Q_Q$ the heavy--quark charge, the light--quark charge matrix $Q$ is
given by
\beq
{\bf Q} \, = \,\left( 
\begin{array}{ccc}
\frac{2}{3} & & \\ & -\frac{1}{3} &  \\ & & -\frac{1}{3} 
\end{array} \right) \ ,
\eeq
and the Goldstone mesons appear through axial--vector, $\xi_{\mu}$, and vector,
$\G_{\mu}$, fields 
\beq
\xi_\mu = \frac{i}{2}\,\left(\xi D_{\mu} \xi^\dagger- \xi^\dagger D_{\mu} \xi
\right) \ , \qquad\qquad
\G_\mu = \frac{1}{2}\,\left(\xi D_{\mu} \xi^\dagger+ \xi^\dagger D_{\mu} \xi
\right) \ ,
\eeq
with $D^{\mu}\xi = \partial^{\mu} \xi-i e {\cal A }^{\mu} [Q , \xi]$.

Because of the different spin configuration of the light degrees of freedom
there is an intrinsic mass difference, $\Delta_{ST}\equiv M_S-M_T$,
between the sextet and triplet baryon multiplets.
Notice that a direct coupling  of the pseudo-Goldstone bosons to the 
$\bar{\bf 3}$ baryons is forbidden at lowest order in $1/\La_\chi$.

The first contributions
to the transitions we are considering  come from:\\

1) the next order ($D=5$) in the baryon chiral Lagrangian \cite{chogeo}
\bea
{\cal L}^{(long)} &=& 
\frac{e}{\Lambda_{\chi}} \left\{ 
i \, c_S \ {\rm tr }\left[\bar{S}_{\mu} Q S_{\nu} +\bar{S}_{\mu} S_{\nu} Q
\right] F^{\mu \nu} 
\right. \nn \\   & & \left.\mbox{}
+ c_{ST}\,\left[\epsilon_{ijk} \,\bar{T}^i v_{\mu} Q_l^j S_{\nu}^{kl} +
\epsilon^{ijk} \,\bar{S}_{\nu,kl} v_{\mu} Q_j^l T_i \right] 
\tilde{F}^{\mu \nu} \right\} \ ,
\label{eq:long}
\eea
where $c_S$ and  $c_{ST}$ are unknown chiral couplings and
$\tilde{F}^{\mu \nu} =\varepsilon^{\mu\nu\a\beta}F_{\a \beta}$ .
We will take $\La_\chi =4\pi f_\pi\simeq 1.2$ GeV, which fixes 
the normalization of these couplings.
A long--distance magnetic moment interaction for just the $T$ baryons does
not exist, since their light quarks are in a $s_l=0$ configuration.

2) terms of order $1/m_{Q}$ in the heavy quark expansion which
break both spin and flavor symmetries \cite{chogeo}
\beq
{\cal L}^{(short)} = 
-\frac{1}{2 m_Q} \,\bar{S}_{ij}^{\lambda} (i D)^2 S_{\lambda}^{ij}
- \frac{e Q_Q}{4 m_Q}\,\bar{S}_{ij}^{\lambda} \sigma_{\mu \nu} S_{\lambda}^{ij}
\, F^{\mu \nu}  
+\frac{1}{2 m_Q} \,\bar{T}^i (i D)^2 T_i 
+ \frac{e Q_Q}{4 m_Q}\,\bar{T}^i \sigma_{\mu \nu} T_i\, F^{\mu \nu}\ ;
\label{eq:hqet}
\eeq

3) chiral loops of Goldstone bosons coupled to photons, as described
by the lowest--order Lagrangian.

\section{Results for $S^*\ra S \g $ decays}
\label{sec:S}

\begin{figure}[ht]
\begin{center}
\epsfig{file=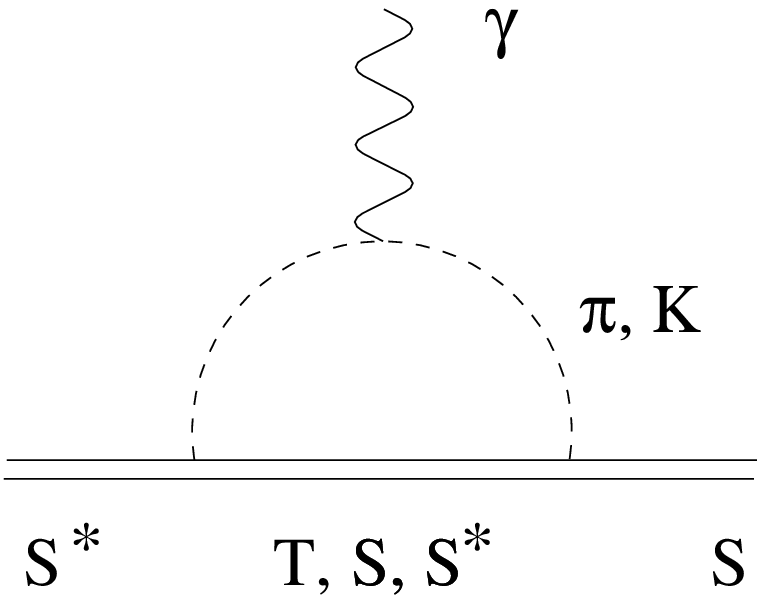,width=0.3\linewidth,angle=0}
\caption{Meson loops contributing to $S^*\ra S \g$.}
\label{fig:S}
\end{center}
\end{figure}

We will decompose our results in two different amplitudes
\beq
{\cal A}\left( B^*\to B \ \g\right) \, = \,
A_{M1}  \, {\cal{O}}_{M1} + A_{E2}  \, {\cal{O}}_{E2} \, ,
\eeq
where the corresponding M1 and E2 operators  are defined by
\bea
{\cal{O}}_{M1}&=& e \, \bar{B}\g_\mu \g_5   B^*_\nu\ F^{\mu\nu},  
\nn \\
{\cal{O}}_{E2}&=& i \, e \, \bar{B}\g_\mu \g_5   B^*_\nu\ v_\a \,
(\partial^\mu F^{\a\nu}+\partial^\nu F^{\a\mu}) \, ,
\label{eq:magop}
\eea
The leading  contributions to M1 transitions come from the
light-- and heavy--quark magnetic interactions  which are of 
 ${\cal O}(1/\La_\chi) $ and  ${\cal O}(1/m_Q) $, respectively.
We have  computed the next-to-leading chiral corrections of  
${\cal O}(1/\La_\chi^2) $,  which originate from the loop diagrams 
shown in fig.~\ref{fig:S}.

\begin{table}
\begin{center} 
\begin{tabular}{|c|c|c|c|c|}  \hl
c quark  & b quark &  $a_\chi$   & $a_{g_2}$ & $a_{g_3}$ 
\\ \hl
$\Sigma^{++*}_{c}\ra\Sigma^{++}_{c}\g $&$\Sigma^{+*}_{b}\ra\Sigma^{+}_{b}\g $
& 2& $I_\pi$ + $I_K$  & $1+m_\pi/m_K$ 
\\
$\Sigma^{+*}_c\ra \Sigma^{+}_c \g$&$\Sigma^{0*}_b \ra \Sigma^{0}_b\g $&
 1/2& $I_K/2$ &    ${1}/{2}$
\\
$\Sigma^{0*}_c \ra \Sigma^{0}_c\g $&$\Sigma^{-*}_b \ra \Sigma^{-}_b\g  $&
$-1$ &$-I_\pi$  & $-m_\pi/m_K$ 
\\
$\Xi^{0'*}_c \ra \Xi^{0'}_c\g $  &$\Xi^{-'*}_b \ra \Xi^{-'}_b\g$  &$-1$&
 $-(I_\pi+I_K)/2$&  $-(1+m_\pi/m_K )/{2}$
\\
$\Xi^{+'*}_c \ra \Xi^{+'}_c \g $  &$\Xi^{0'*}_b\ra \Xi^{0'}_b \g $  & 1/2&
$I_\pi/2$ &  ${m_\pi}/(2 m_K)$
\\
$\Omega^{0*}_c\ra \Omega^{0}_c \g $ &$\Omega^{-*}_b\ra \Omega^{-}_b\g $ &$-1$
&  $-I_K$ &  $ -1$
\\ \hl
\end{tabular} 
\caption{Contributions to  M1 amplitudes for $S^* \ra S \g$. }
\label{tab:uno}
\end{center} 
\end{table}

The resulting M1 amplitudes  can be written as:
\beq
A_{M1} (B^{*}) = \fr{1}{\sqrt{3}} \left( - \fr{Q_{Q}}{m_Q} -
\fr{2 c_s}{3 \Lambda_\chi}\, a_{\chi}(B^{*}) 
+ g_2^{2} \,\fr{\Delta_{ST}}{4 (4 \pi f_\pi)^2}\, a_{g_2}(B^{*})+
g_3^{2} \,\fr{m_K}{4 \pi f_\pi^2}a_{g_3}\, (B^{*}) \right) \ .
\label{eq:magS}
\eeq
In Table~\ref{tab:uno} we show the values of the coefficients $a_i(B^*)$ 
for the decays of baryons containing one charm or bottom quark.
In the  table,
\beq
\label{eq:intr}
I_i\equiv I(\Delta_{ST}, m_i) = 2 \left(-2 +\log{\fr{m_i^{2}}{\mu^{2}}}
\right)+ 2 \, {\sqrt{\Delta_{ST}^{2}- m_i^{2}}\over \Delta_{ST}}\,
\log{\left(\fr{\Delta_{ST}+ \sqrt{\Delta_{ST}^{2}- m_i^{2}} }{ \Delta_{ST}- 
\sqrt{\Delta_{ST}^{2}- m_i^{2}}
 }\right)} \ .
\eeq

Due to flavor symmetry,  all contributions are equal 
for charm and bottom baryons, with the only exception of 
the term proportional to  the heavy  quark electric charge
($Q_{c}=+2/3$, $Q_{b}=-1/3$).
The  calculation of the decay amplitudes closely follows the one reported
in ref.~\cite{nos} for the magnetic moments of the $S^{(*)}$ baryons.
Thus,  we list  the  arguments  common to both calculations:

\begin{enumerate}
\item   
contributions of ${\cal O}(1/(m_{Q}\Lambda_{\chi}))$
can be neglected for the $b$ baryons. For the $c$ baryons, 
they are expected to be smaller than $15\%$ \cite{nos}.

\item
the  corrections  proportional to $g_2^{2}$ are obtained performing a 
one--loop integral (fig.~\ref{fig:S} with an $S$ baryon running in the loop) 
that has to be renormalized.
The divergent part of the integral does not depend on the pion or kaon masses
and is instead proportional to the  mass of the baryon running in the loop.
If one considers both pion and kaon loops the divergent part respects
the $SU(3)$ structure of the  chiral multiplet and  can be canceled
with an operator of the  form
\beq
i \frac{e}{\Lambda_{\chi}^2}\,
{\rm tr }\left[\bar{S}_{\mu}\left(v\cdot D S_{\nu}\right) Q
-\left( v\cdot D\bar{S}_{\mu}\right) S_{\nu}\, Q
\right] F^{\mu \nu}  \ . 
\label{eq:dseis}
\eeq
This is the most general dimension--6 chiral-- and Lorentz--invariant
operator, constructed out from $S_\mu^{ij}$ and $Q\, F^{\mu\nu}$,
preserving parity and time--reversal invariance,
which contributes to the M1 amplitudes.
When the equation of motion 
$i\, (v\cdot D)\, S_{\mu}\,=\, \Delta_{ST}\, S_{\mu}$ is applied,
its contribution is of the same form as the term proportional to $c_{s}$ in
 \\ Eq.~(\ref{eq:long}).
Thus, the local contribution from the operator in Eq.~(\ref{eq:dseis})
can be taken into account, together 
with the lowest--order term in Eq.~(\ref{eq:long}),
through an effective coupling $c_S(\mu)$.
The scale $\mu$ dependence of the loop integrals is exactly canceled
by the corresponding dependence of the coefficient $c_S(\mu)$;

\item
the contribution proportional to $g_3^{2}$ involves  a loop integral
with a baryon of the $T$ multiplet running in the loop.
Since the Lagrangian does not have any mass term for $T$ baryons,
the result of the integral is convergent and proportional to  
the mass of the light mesons.
\end{enumerate}

\begin{figure}[ht]
\begin{center}
\epsfig{file=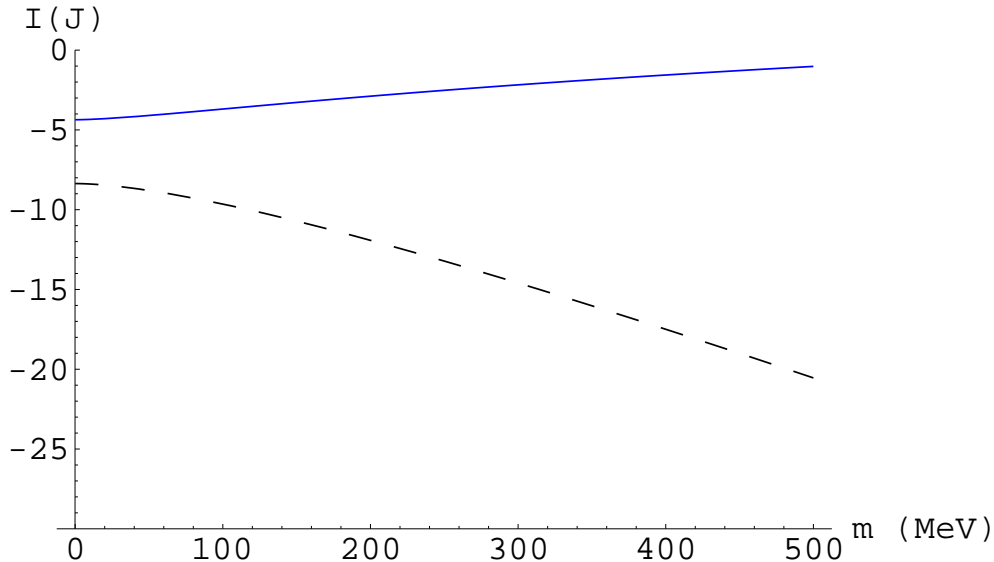,width=0.8\linewidth,angle=0}
\caption{ The scaling  of the functions $I(\Delta_{ST},m)$,
 Eq.~(\ref{eq:intr}), 
and $J(\Delta_{ST},m)$,
Eq.~(\ref{eq:jj}) as a function of the meson mass $m$. The  dashed line is
 $I(\Delta_{ST},m)$  and the continuous line is $J(\Delta_{ST},m)$.
The scale $\mu$ is fixed at 1 GeV and $\Delta_{ST}=168$ MeV.}
\label{fig:ir}
\end{center}
\end{figure}

In order  to see the  behavior of $I(\Delta_{ST}, m)$
   with the meson mass we have plotted
 it  in fig.~\ref{fig:ir},  for $\mu=1$ GeV and $\Delta_{ST}$ as in 
Table~\ref{tab:cost}.
 We see that  the value of $I(\Delta_{ST},m)$  raises considerably
 in the limit of zero Goldstone-boson mass.

\begin{table}[htb]
\begin{center}
\begin{tabular}{|c|c|} \hl
$f_\pi$ & 93 MeV \\
$m_\pi$ & 140 MeV \\
$m_K$ & 496.7 MeV \\
$\Delta_{ST}$ & 168 MeV \\
$m_c$ & 1.3 GeV \\
$\a_{em}(m_\tau)$&1/133.3\\ 
$m_b$ & 4.8 GeV \\ \hl
\end{tabular}
\caption{Constants used in  numerical estimates.}
\label{tab:cost}
\end{center}
\end{table} 

From Table~\ref{tab:uno}, one  can derive  the following
linearly--independent relations for  the M1 amplitudes of the $S^{*}\ra S\g $
decays containing a charm quark:
\bea
A_{M1}(\S^{++*}_c)&=&2A_{M1}(\S^{+*}_c)- A_{M1}(\S^{0*}_c)  =
2 A_{M1}(\Xi^{+'*}_c)-A_{M1}(\Omega^{0*}_c) 
\nn \\
A_{M1}(\S^{++*}_c)+2A_{M1}(\Xi^{0'*}_c)
&=& A_{M1}(\S^{0*}_c)+2A_{M1}(\Xi^{+'*}_c)   = \,
 -\fr{2}{\sqrt{3} m_c}\ . 
\label{eq:prel}
\eea
The ${\cal O}(1/\Lambda_{\chi})$ and ${\cal O}(1/\Lambda_{\chi}^2)$
contributions  cancel in the sum of the six $S^*\to S\gamma $ M1
amplitudes. Therefore, the average 
over the baryon sextet measures the $O(1/m_Q)$ contribution.

We can write four analogous relations for the bottom baryons:
\bea
A_{M1}(\S^{+*}_b)&=&2A_{M1}(\S^{0*}_b)-A_{M1}(\S^{-*}_b)  
= 2 A_{M1}(\Xi^{0'*}_b)-A_{M1}(\Omega^{-*}_b)  
\nn \\
A_{M1}(\S^{+*}_b)+2A_{M1}(\Xi^{-'*}_b)
&=&A_{M1}(\S^{-*}_b)+2A_{M1}(\Xi^{0'*}_b)
 = \, \fr{1}{\sqrt{3} m_b}  
\label{eq:prela} 
\eea
Two additional equations  relate $b$ and $c$ baryons:
\bea
A(\S^{+*}_c)-A(\S^{++*}_c)&=&A(\S^{0*}_b)-A(\S^{+*}_b)
\nn \\
A(\S^{+*}_c)-A(\Xi^{+'*}_{c })&=& A(\S^{0*}_b)-A(\Xi^{0'*}_{b }) \ .
\label{eq:prelb}
\eea

The same diagram in fig.~\ref{fig:S}
generates the leading  contributions to E2 transitions.
The graph is of ${\cal O}(1/\La_\chi^2)$ and  
one has to include all chiral counterterms  up to this order.
There is only one operator  with these features,
\beq
\fr{i}{4}
{e\, c^{E2}_S\over\La_\chi^2}\,
{\rm tr }\left[\bar{S}_{\mu} Q S_{\nu} +\bar{S}_{\mu} S_{\nu} Q
\right] v_\alpha \,\left(\partial^\mu F^{\alpha \nu} 
+ \partial^\nu F^{\alpha \mu} \right) \, , 
\eeq
and so only one  new unknown constant ($c^{E2}_S$) appears.
The E2 amplitudes can be written analogously to the M1 case:
\bea
A_{E2} (B^{*})&=&\fr{1}{6\sqrt{3}} \left(  
\fr{ c^{E2}_S}{ \Lambda_\chi^2}\ b_{\chi}(B^{*})-
\fr{g_2^{2}}{4 (4 \pi f_\pi)^2} \, b_{g_2}(B^{*})-
\fr{g_3^{2}}{4 \pi f_\pi^2} \, b_{g_3}(B^{*}) \right)\ .
\label{eq:magSE}
\eea
The coefficients $b_i$ are shown in table~\ref{tab:unoe}, where 
\bea
J_i&\equiv&J(\Delta_{ST},m_i)=\fr{\partial}{ \partial \Delta_{ST}} 
\left[ \Delta_{ST}I(\Delta_{ST}, m_i) \right] \ ,
\nn \\
J_i^0&=&\lim_{\Delta \ra 0} J_i=-1-i\pi + \log{m_i/\mu} \ .
\label{eq:jj}
\eea
The scale dependence  of $c^{E2}_S(\mu)$ cancels the one coming from the loop
calculation.
While  the behavior of $J(\Delta_{ST},m_i)$ does not  change  much
when one varies  the meson mass (see fig.~\ref{fig:ir}), $J_i^0$
  is  infrared divergent in the exact chiral limit.
This divergence can be responsible for  a considerable enhancement
 of the electric dipole effects.

\begin{table}
\begin{center} 
\begin{tabular}{|c|c|c|c|c|}  \hl
 c quark  & b quark & $b_\chi $ & $b_{g_2} $ & $b_{g_3} $ 
\\ \hl
$\Sigma^{++*}_{c}\ra\Sigma^{++}_{c}\g $&$\Sigma^{+*}_{b}\ra\Sigma^{+}_{b}\g $
& 2&$J_\pi$ + $J_K$ &$J_\pi^0$ + $J_K^0$
\\
$\Sigma^{+*}_c\ra \Sigma^{+}_c \g$&$\Sigma^{0*}_b \ra \Sigma^{0}_b\g $&
 1/2&  $J_K/2$ &  $J_K^0/2$ 
\\
$\Sigma^{0*}_c \ra \Sigma^{0}_c\g $&$\Sigma^{-*}_b \ra \Sigma^{-}_b\g  $&
$-1$ &$-J_\pi$   &$-J_\pi^0$ 
\\
$\Xi^{0'*}_c \ra \Xi^{0'}_c\g $  &$\Xi^{-'*}_b \ra \Xi^{-'}_b\g$  &$-1$&
$-(J_\pi+J_K)/2$ & $-(J_\pi^0+J_K^0)/2$ 
\\
$\Xi^{+'*}_c \ra \Xi^{+'}_c \g $  &$\Xi^{0'*}_b\ra \Xi^{0'}_b \g $  & 1/2&
$J_\pi/2$ &$J_\pi^0/2$  
\\
$\Omega^{0*}_c\ra \Omega^{0}_c \g $ &$\Omega^{-*}_b\ra \Omega^{-}_b\g $ 
&$-1$&  $ -J_K$ &  $-J_K^0$
\\ \hl
\end{tabular} 
\caption{Contributions to E2 amplitudes for $S^* \ra S \g$. }
\label{tab:unoe}
\end{center} 
\end{table}

The M1 and E2 amplitudes have identical $SU(3)$ structure. Therefore,
we can construct for the  E2 amplitudes 
exactly the same relations as in the M1 case 
[Eq.~(\ref{eq:prel}-\ref{eq:prelb})]. 
However as there are no $1/m_Q$ terms contributing to E2, the last
equations in~(\ref{eq:prel}) and~(\ref{eq:prela}) must be replaced by
\beq
A_{E2}(\S^{++*}_c)+2A_{E2}(\Xi^{0'*}_c) = 
A_{E2}(\S^{0*}_c)+2A_{E2}(\Xi^{+'*}_c)= 0 \ ,
\eeq
and
\beq
A_{E2}(\S^{+*}_b)+2A_{E2}(\Xi^{-'*}_b) = 
A_{E2}(\S^{-*}_b)+2A_{E2}(\Xi^{0'*}_b)=0 \ .
\eeq

The electromagnetic decay widths  are given by
\beq
\G (S^*\ra S \g )=\fr{4 \a_{em}}{3}\,\fr{E_\g^3 M_S}{ M_{S^*}}\,
\left( |A_{M1}|^2+3E_\g^2 |A_{E2}|^2 \right) \ ,
\label{eq:width}
\eeq
where $ M_{S^*}$ and $M_S$ are the masses of the initial and final baryons
and $E_\g$ the energy of the outgoing photon.

The  E2 amplitudes  come at higher chiral order with respect to the M1 ones.
Therefore, the E2 contribution to the total width is suppressed  by  a 
factor $(E_\g/\La_\chi)^2\sim 5\%$.
In principle, it should be possible to determine experimentally  the
ratio $A_{E2}/A_{M1}$ by studying the angular distribution of photons
from the decay of polarized baryons \cite{savanu,RB:67,BSS:93}.
The Fermilab E-791 experiment has reported \cite{E791}
a significant polarization effect on the production of $\Lambda_c$ baryons,
which perhaps could be useful in future measurements of these
electromagnetic decays.

In order to provide an absolute theoretical prediction for all the 
decay widths, it is necessary to have an estimate of the
couplings $c_S$, $g_2$ and $g_3$ (we neglect for the moment
the small E2 contamination).
The couplings $g_2$ and  $g_3$ have been calculated theoretically
\cite{gnuc,gural,grozi,zhu}; we report the results of these computations 
in Table~\ref{tab:gteor}.

There exists  an  experimental measurement of $g_3$  from CLEO
coming from the  decay $\S_c^*\rightarrow \Lambda_c \pi $~\cite{cleo,gnec},
$g_3= 
0.99\pm 0.17$.
The direct measurement of $g_2$ is not possible at present.
However, the quark model relates its value to $g_3$~\cite{gnec}, yielding
$g_2=1.40 \pm 0.25$.

\begin{table}[htb]
\begin{center}
\begin{tabular}{|c|c|c|} \hl
Model &$g_2$ & $g_3$ \\ \hl
Large $N_c$ ~\cite{gural} & 1.88 & 1.53\\
Quark model ~\cite{gnuc} & $1.5$ & $1.06 $ \\
Short--distance QCD sum rule ~\cite{grozi} & $0.83 \pm 0.23$ & $0.67 \pm 0.18$\\
Light--cone  QCD sum rules ~\cite{zhu} & $1.56\pm 0.3\pm 0.3$ &
$0.94 \pm 0.06\pm 0.2$ \\ \hl
\end{tabular}
\caption{Theoretical estimates of $g_2$ and $g_3$.}
\label{tab:gteor}
\end{center}
\end{table} 

The constant $c_S$ is {\it a priori} unknown and its value should be 
extracted from the experiment or predicted by some more fundamental model.
This coupling appears also in the calculation of the magnetic moments of
$S^{(*)}$  baryons~\cite{nos}.
Thus, the determination of its numerical value via the measurement of any
of these electromagnetic decays, would also provide an absolute prediction 
for the magnetic moments.

Having a numerical determination of the couplings $g_2$ and $g_3$,
it is possible to derive  a scale independent relation  between any couple 
of M1 (E2) amplitudes. The combinations 
\beq
A_{M1}(B_1^*) - \fr{a_\chi(B_1^*)}{a_\chi(B_2^*)} \ A_{M1}(B_2^*), \quad
A_{E2}(B_1^*) - \fr{b_\chi(B_1^*)}{b_\chi(B_2^*)} \ A_{E2}(B_2^*)
\eeq
are independent of the unknown coupling $c_S(\mu)$  and can then be predicted.
For instance
\beq
A_{M1}(\S^{++*}_c)+2 \, A_{M1}(\S^{0*}_c) = \frac{1}{\sqrt{3}}\,
\fr{g_3^{2}}{4 \pi f_\pi^{2}} \, (m_K-m_\pi)+
\fr{\Delta_{ST}}{4 \sqrt{3}}\,\fr{g_2^{2}}{(4 \pi f_\pi)^{2}} \,
(I_K-I_\pi) - \fr{2}{\sqrt{3} m_c}\ .
\label{eq:dif}
\eeq
In order to get a numerical estimate of the left--hand side of 
Eq.~(\ref{eq:dif})
we set $g_2=1.5 \pm 0.3$, $g_3=0.99 \pm 0.17$ 
and the rest of the constants as in Table~\ref{tab:cost}.
We find then 
%
\beq
A_{M1}(\S^{++*}_c)+2 A_{M1}(\S^{0*}_c)
= 0.57 \pm 0.67 \ {\rm GeV^{-1}}
\label{eq:aul}
\eeq
The  analogous relation for $b$ baryons reads
\beq
A_{M1}(\S^{+*}_b)+2 A_{M1}(\S^{-*}_b)
= 1.58 \pm 0.66 \ {\rm GeV^{-1}}
\label{eq:aulb}
\eeq

The main contribution to these values corresponds to the chiral loop, 
with a much smaller correction coming from the $1/m_Q$ term. 
These sums would be zero if none of the previous contributions were included.
The large  errors  in Eq.~(\ref{eq:aul}) and Eq.~(\ref{eq:aulb}) come  from
the  present  uncertainties on $g_{2,3}$ ($\sim 20\% $).
The   same consideration holds  for all numerical results
in this and in the following sections.

 A further comment is now in order.
To estimate the importance
 of the effect  of  one loop HHCPT we define  the ratio 
(see Eq.~(\ref{eq:magS})),
\beq 
R(B^*)=\fr{3 }{ 2 (4\pi f_\pi) }\fr{g_2^2 \ \Delta_{ST}\ a_{g2}(B^*)/4\ +
4\pi\  m_K \  g_3^2 \ a_{g3}(B^*) }{|c_S(\mu)|\  a_\chi(B^*)  } \ .
\eeq
 We find for $\La_\chi/2< \mu < \La_\chi $,
\bea
R(\S^{++*}_{c})=R(\S^{+*}_{b})=& R(\Xi^{0'*}_{c})=R(\Xi^{-'*}_{b})&=
\fr{3.2\pm 1.9}{|c_S(\mu)|}, \nn  \\
R(\S^{+*}_{c})=R(\S^{0*}_{b})= & R(\Omega^{0*}_{c})= R(\Omega^{-*}_{b})&=
 \fr{5.5 \pm 2.9}{|c_S(\mu)|},\nn \\
R(\S^{0*}_{c})=R(\S^{-*}_{b})=& R(\Xi^{+'*}_{c})=R(\Xi^{0'*}_{b})&=
 \fr{1.0\pm 1.1}{|c_S(\mu)|} \ .
\label{eq:rat}
\eea
The  scale dependence of this result is not very strong  and  in any case
 within the errors.
Na\"\i vely one expects $|c_S(\mu)|\sim {\cal O}(1)$.
Thus, from Eq.~(\ref{eq:rat}), we can deduce that 
   the   infrared effect due to the coupling of the photon to light
mesons, is large  on  these electromagnetic decays.
This affirmation can be sustained also comparing our results with some
estimates existing in the literature
(so far there are no experimental data on $S^* \ra S \g$ decays).
In ref.~\cite{redai} the three decays $\S^{*}_b \ra \S_b \g$
are predicted, using light cone QCD sum rules;
these results respect the HQET and chiral symmetries and
 agree with the first of our relations in Eq.~(\ref{eq:prela}),
provided the proper relative signs among the amplitudes are chosen, namely
\beq
\fr{\sqrt{\G(\S^{*+}_b \ra \S^+_b \g)}-\sqrt{\G(\S^{*-}_b \ra \S^-_b \g)}}{
2 \sqrt{\G(\S^{*0}_b \ra \S^0_b \g)}} = 0.98 \ .
\label{eq:chino}
\eeq
In order to
 derive this number from the results of ref.~\cite{redai}, we have made use
of the baryon masses in Table~\ref{tab:mass}.
However  in  ref.~\cite{redai} all coupling constants are determined 
 at leading order in HQET.
   Writing
\beq
c_S(\mu)_{\overline{MS}}=c_S^0 + \fr{
\left(c_S^1(\mu)\right)_{\overline{MS}}}{\La_\chi}
\eeq
we derive (consistently with Eq.~(\ref{eq:chino})) from ref.~\cite{redai}
\bea
-1.6<c_S^0<-1.2 , & \ {\rm or }, \  & 1.3<c_S^0<1.7
\label{eq:cszero}
\eea
depending on the overall sign of the 
amplitudes\footnote{The difference in the absolute value of positive
 and negative interval in Eq.~(\ref{eq:cszero}) 
is due to the heavy quark term in Eq.~(\ref{eq:magS}).}.
Thus   ref.~\cite{redai}, obtains the  expected  order of 
 magnitude of $c_S^0$, however, the  important  chiral effect
due to  the photon--meson  coupling
  is not taken into account .
Thus,  choosing  the sign  between the amplitudes consistently with
Eq.~(\ref{eq:chino}),  it is impossible to deduce  Eq.~(\ref{eq:aulb})
 from their calculation.
Ref.~\cite{chori} estimates these same  decay rates and its results 
are consistent with ref.~\cite{redai} so that   the same  comments
are valid  also for this reference.
These  considerations  apply
 also if one considers the computation
 of the decays $\S^{*}_c \ra \S_c \g$ of ref.~\cite{gnec}.
In this case the first of our  relations in Eq.~(\ref{eq:prel}) is exactly
fulfilled and  we can derive $|c_S^0|=1\pm 1$.
We note however  that 
 the predictions of ref.~\cite{chori,ivan} and ref.~\cite{gnec}
 for $\G(\S^{*+}_c \ra \S_c^+ \g)$ are in desagreement as  a much higher
 rate is predicted in the first two references.

\begin{table}
\begin{center} 
\begin{tabular}{|c|c|c|c|}  \hl
 $c$-baryons  &M (MeV) & $b$-baryons  & M (MeV) \\
\hl
$\Xi^{0}_c$ & $2470.3 \pm 1.8 $& $\Xi^{-}_b$ & $5805.7 \pm 8.1 $ \\
$\Xi^{+}_c$ & $2465.6 \pm 1.4$& $\Xi^{0}_b$ & $5805.7 \pm 8.1 $ \\
$\La_c^{+}$ & $2284.9 \pm 0.6$& $\La_b^{0}$ & $5624 \pm 9$ \\
$\S^{++}_c$ & $2452.8 \pm 0.6$& $ \S^{+}_b$ & $5824.2\pm 9.0 $ \\
$\S^{+ }_c$ & $2453.6 \pm 0.9$& $ \S^{0}_b$ & $5824.2\pm 9.0 $ \\
$\S^{0 }_c$ & $2452.2 \pm 0.6$& $ \S^{-}_b$ & $5824.2\pm 9.0 $ \\
$\Xi^{0'}_c$ &$2577.3 \pm 3.2$& $\Xi^{-'}_b$ & $5950.9\pm 8.5 $ \\
$\Xi^{+'}_c$ &$2573.4 \pm 3.1$& $\Xi^{0'}_b$ & $5950.9\pm 8.5 $ \\
$\Omega^{0}_c$&$2704.0 \pm 4.0$& $\Omega^{-}_b$&$6068.7\pm 11.1$\\
$\S^{++*}_c$ & $2519.4 \pm 1.5$& $ \S^{+*}_b$ & $5840.0 \pm 8.8 $ \\
$\S^{+ *}_c$ & $2518.6 \pm 2.2$& $ \S^{0*}_b$ & $5840.0 \pm 8.8 $ \\
$\S^{0 *}_c$ & $2517.5 \pm 1.4$& $ \S^{-*}_b$ & $5840.0 \pm 8.8$ \\
$\Xi^{0'*}_c$ &$2643.8 \pm 1.8$& $\Xi^{-'*}_b$ & $5966.1 \pm 8.3 $ \\
$\Xi^{+'*}_c$ &$2644.6 \pm 2.1$& $\Xi^{0'*}_b$ & $5966.1 \pm 8.3 $ \\
$\Omega^{0*}_c$&$2760.5 \pm 4.9$& $\Omega^{-*}_b$&$6083.2 \pm 11.0$\\
\hl
\end{tabular} 
\caption{Masses of charm and bottom baryons. All masses of $b$ baryons 
(except $\La_b^{0}$)  and
the ones of $\S^{+ *}_c$, $\Omega^{0*}_c$ have been estimated theoretically 
in  ref.~\protect\cite{pola}. The measured masses are taken 
from~\protect\cite{caso}.
}
\label{tab:mass}
\end{center} 
\end{table}

\section{Results for $S^* \ra T \g$ decays}
\label{sec:T}

The M1 and E2 operators for these decays are defined as in Eq.~(\ref{eq:magop}).
Similarly to what we have done in the previous paragraph, we write
the M1 amplitude for $S^* \ra T \g$ decays as
\beq
A_{M1} (B^*) = 
 -\sqrt{2}\, \fr{ c_{ST}}{ \Lambda_\chi } \, a_{\chi}(B^*)+
 g_2\, g_3 \,\fr{\Delta_{ST}}{2 \sqrt{2} (4 \pi f_\pi)^2}\, a_{g}(B^*) \ .
\label{eq:amt}
\eeq
The values  of  the  coefficients $a_i$ are written in Table~\ref{tab:t}.

\begin{table}
\begin{center} 
\begin{tabular}{|c|c|c|c|}  \hl
c quark & b quark & $ a_{\chi}  $   & $a_g   $ \\ \hl
$\S^{+*}_c \ra \Lambda_c^{+} \g$   &$\S^{0*}_b \ra \Lambda_b^{0} \g$   &$1$&
$2 I_{\pi}+I_K/2$\\
$\Xi^{+'*}_c \ra \Xi^{+}_c \g$  &$\Xi^{0'*}_b \ra \Xi^{0}_b \g $  &$1$&
$I_{\pi}/2+2 I_K$ \\
$\Xi^{0'*}_c \ra \Xi^{0}_c \g$  &$\Xi^{-'*}_b \ra \Xi^{-}_b \g $  &$0$ &
$-I_\pi/2+I_K/2$ \\ \hl
\end{tabular} 
\caption{Contributions to M1 amplitudes for $S^* \ra T\gamma$.  }
\label{tab:t}
\end{center} 
\end{table}

\begin{figure}[ht]
\begin{center}
\epsfig{file=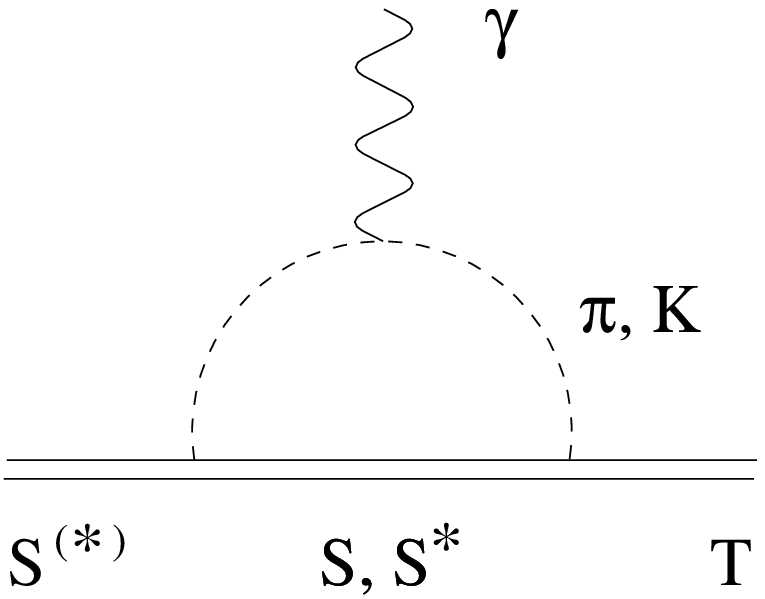,width=0.3\linewidth,angle=0}
\caption{Meson loops contributing to $S^{(*)}\ra T \g$.}
\label{fig:T}
\end{center}
\end{figure}

The first term in Eq.~(\ref{eq:amt}) comes from the Lagrangian~(\ref{eq:long}),
while the second one corresponds to the diagram of fig.~\ref{fig:T}.
As in the case of $S^* \ra S \g$, when all Goldstone boson loops are included,
the scale $\mu$ dependence of the loop diagram 
is canceled by  the   corresponding  dependence  of an effective  
$c_{ST}(\mu)$.
After applying the equations of motion,
the effective coupling $c_{ST}(\mu)$ contains all contributions to 
the M1 amplitude 
coming from ${\cal O}(1/\Lambda_{\chi}^2)$ counterterms, namely
\bea
& i \,\epsilon_{ijk}\, Q_l^j \,\left( \bar{T}^i (v \cdot D  S_{\nu}^{kl}) - 
(v \cdot D \bar{S}_{\nu}^{kl})  T^i \right) v_{\mu} \tilde{F}^{\mu \nu} & \ ,
\nn \\
&i \,\epsilon_{ijk}\, Q_l^j\,  \left(\bar{T}^i  (D_{\mu}  S_{\nu}^{kl})- 
(D_{\mu} \bar{S}_{\nu}^{kl})  T^i \right) \tilde{F}^{\mu \nu} & \ .
\eea

Our result in Eq.~(\ref{eq:amt}) does not depend on the heavy quark 
charge or mass.
We thus obtain the same predictions for charm and bottom baryons.
All constants  can be eliminated in the relations 
\bea
A_{M1}(\S^{+*}_c)-A_{M1}(\Xi^{+'*}_c)& =&  -3\, A_{M1}(\Xi^{0'*}_c)\ , \nn \\
A_{M1}(\S^{0*}_b)-A_{M1}(\Xi^{0'*}_b)& =&  -3\, A_{M1}(\Xi^{-'*}_b) \ .
\label{eq:innom}
\eea

It is interesting to notice that $A_{M1}(\Xi^{0'*}_c)$ does not depend 
on $c_{ST}$.
Since at ${\cal O}(1/\Lambda_{\chi}^2)$
this decay does not get any contribution from local terms,
its M1 amplitude results from a {\it finite} chiral loop calculation
(it cannot be divergent because there is no possible counter-term to
renormalize it), so that we have an absolute prediction for its value
in terms of $g_2$ and $g_3$.
Using for $g_2$ and $g_3$ the same values as in Eq.~(\ref{eq:aul}),
we find
\beq
\G_{M1}(\Xi^{0'*}_c)=5.1 \pm 2.7 \ {\rm KeV} \ .
\label{eq:xic}
\eeq
Lower estimates of this decay width are reported in ref.~\cite{ivan}, 
$\G_{M1}(\Xi^{0'*}_c)=0.68 \pm 0.04\ {\rm KeV}$, and in ref.~\cite{gnec},
$\G_{M1}(\Xi^{0'*}_c)=1.1\ {\rm KeV} $.
These authors do not consider chiral corrections to their result which
 cannot be neglected.
 In particular the result  of
 ref.~\cite{gnec}  is worth a further comment.
The effective coupling
 to the M1 operator  in this decay is estimated  using the non relativistic
quark model. This coupling is found to be  proportional to $1/M_d-1/M_s$, where
$M_{d,s}$ are the constituent quark mass of the  down and strange  quarks respectively.
However
\beq
\fr{1}{M_d}-\fr{1}{M_s}=\fr{M_s-M_d}{M_s M_d}\sim
 {\cal O}\left(\fr{m_K^2-m_\pi^2}{\Lambda_\chi^3}\right) \ .
\eeq
Thus,
 the effect they calculate  represents a higher order correction to our result.

The corresponding decay for $b$ baryons, $\Xi^{-'*}_b\ra \Xi^{-}_b\gamma$
can be also predicted, using the existing estimates for the masses of these 
baryons (see Table~\ref{tab:mass}),
\beq
\G_{M1}(\Xi^{-'*}_b)=4.2 \pm 2.4 \ {\rm KeV} \ .
\label{eq:xib}
\eeq
The dominant error of Eq.~(\ref{eq:xic}) and Eq.~(\ref{eq:xib})
 comes from the determination of the 
couplings $g_{2,3}$.

The E2 amplitude in  $S^*\to T\gamma$ 
is suppressed by an extra power of $1/m_Q$.
The first non-zero contributions comes at  
${\cal O}(1/m_Q \Lambda_{\chi}^2)$.
At this order we find:

\begin{itemize}

\item
a divergent contribution~\cite{savanu} arising from the lowest--order 
Lagrangian~(\ref{eq:lagos}), through the loop in fig.~\ref{fig:T},
which is proportional to the mass splitting between 
$S$ and $S^*$ baryons~\cite{jenk},
\beq
\Delta M_Q =3 \fr{\lambda_{2 S}}{m_Q} \ ;
\eeq
\item
a spin symmetry breaking operator of ${\cal O} (1/m_Q)$,
\beq
{\cal{L}}^{\prime} = i\, \fr{g'}{m_Q}\,
\left [\epsilon_{ijk} \,\bar{T}^i\sigma^{\mu\nu} (\xi_{\mu})_l^j S_{\nu}^{kl}+
\epsilon^{ijk} \,\bar{S}_{kl}^{\mu} \sigma_{\mu\nu} (\xi^{\nu})_j^l T_i\right ] \ ,
\label{eq:nova}
\eeq
which gives rise to divergent loop diagrams, as the one in fig.~\ref{fig:T},
 where one
of the vertices is proportional to $g'$;

\item
further, there are finite contributions of the same order coming from
\beq
-i \fr{c^{E2}_T}{m_Q \Lambda_{\chi}^2} \,\eps_{ijk}\,
\bar{T}^i\sigma_{\mu\nu} Q_l^j S_{\a}^{kl} 
\,\partial^{\a} \tilde{F}^{\mu \nu} \ .
\label{eq:counte}
\eeq
We could also include the operator 
\beq
i \eps_{ijk}\,\bar{T}^i\sigma_{\mu\nu} Q_l^j S_{\a}^{kl} 
\,\partial^{\nu} \tilde{F^{\mu \a}} \ ,
\eeq
but its contribution is proportional to that in Eq.~(\ref{eq:counte})
up to higher order corrections.

\end{itemize}

\begin{table}
\begin{center} 
\begin{tabular}{|c|c|c|c|c|}  \hl
c quark & b quark &  $ b_{\chi}  $   & $b_{g'} $ & $b_g$ \\ \hl
$\S^{+*}_c \ra \Lambda_c^{+} \g$   &$\S^{0*}_b \ra \Lambda_b^{0} \g$   &$1$&
$2 J_{\pi}+J_K/2$ & $2 G_{\pi}+G_K/2$
\\
$\Xi^{+*}_c \ra \Xi^{+}_c \g$  &$\Xi^{0*}_b \ra \Xi^{0}_b \g $  &$1$&
$J_{\pi}/2+2 J_K$ & $G_{\pi}/2+2 G_K$ 
\\
$\Xi^{0*}_c \ra \Xi^{0}_c \g$  &$\Xi^{-*}_b \ra \Xi^{-}_b \g $  &$0$ &
$-J_\pi/2+J_K/2$ & $-G_\pi/2+G_K/2$ 
\\ \hl
\end{tabular} 
\caption{Contributions to E2 amplitudes for $S^* \ra T \gamma$.  }
\label{tab:te}
\end{center} 
\end{table}

Finally, the E2 amplitude can be written as
\beq
A_{E2}(B^*)=-\fr{1}{\sqrt{2}} \fr{c^{E2}_T}{m_Q \Lambda_{\chi}^2}\, b_{\chi}(B^*)
- \fr{1}{24 \sqrt{2}} \fr{g' g_2}{m_Q (4 \pi f_{\pi})^2}\, b_{g'}(B^*)
+\fr{\la_{2S}}{24 \sqrt{2}} \fr{g_2 g_3}{m_Q (4 \pi f_{\pi})^2}\, b_{g}(B^*) \ .
\eeq
The values of the different contributions are collected in Table~\ref{tab:te},
where
\beq
G_i= \left. \fr{\partial\  J_i}{ \partial \Delta_{ST}} 
 \right|_{\Delta_{ST}=0}=\fr{2 \pi}{m_i}\ .
\eeq
We  underline the infrared divergent  behavior of this term.
Neither the  interaction  in Eq.~(\ref{eq:nova})
nor the local term Eq.~(\ref{eq:counte}) 
have been taken into account in the literature. 
An estimate of E2 for $\S^{+*}_c \ra \Lambda^+_c \g $
is provided in ref.~\cite{savanu}, considering only the 
$b_g$ contribution.

By eliminating the unknown coupling constants,
one  can deduce the relation
\beq
A_{E2}(\S^{+*}_c)-A_{E2}(\Xi^{+'*}_c) =  -3 \, A_{E2}(\Xi^{0'*}_c)
\ .
\eeq
The same relation holds for the corresponding $b$ baryons, since
\beq
A_{E2}(B^*_b)=\fr{m_c}{m_b} \, A_{E2}(B^*_c) \ .
\eeq

The decays $\Xi^{0*}_c \ra \Xi^{0}_c \g$   and $\Xi^{-*}_b \ra \Xi^{-}_b \g $
do not get any contribution from the local term proportional to
$c^{E2}_T$; their $O(1/m_Q\Lambda_\chi^2)$ E2 amplitude is
also given by a finite loop calculation. Unfortunately,
since the coupling $g'$ is not known, there is no absolute prediction
in this case.
An experimental measurement of these E2 amplitudes 
would provide a direct estimate of $g'$.

\section{Results for $S \ra T \g$}
\label{sec:st}

The calculation of the M1 amplitude for $S \ra T \g$ decays is analogous 
to that of the previous section.
Now the M1 operator is defined as
\beq
{\cal O}_{M1}=i e \,\bar{B}_T \sigma_{\mu \nu} B_S \, F^{\mu \nu} 
\eeq
and the corresponding amplitude can be written in the form
\beq
A_{M1} (B) = 
\fr{1}{\sqrt{6}} \fr{ c_{ST}}{ \Lambda_\chi }\, a_{\chi}(B)-
g_2 g_3 \,\fr{\Delta_{ST}}{4 \sqrt{6} (4 \pi f_\pi)^2}\, a_{g}(B) \ ,
\label{eq:ams}
\eeq
where the coefficients satisfy
\beq
a_{\chi}(B)=a_{\chi}(B^*) , \qquad\quad a_{g}(B)=a_{g}(B^*) \ .
\eeq
Therefore, the relation~(\ref{eq:innom}) is also valid in this case.
The widths of the decays
$\Xi^{0'}_c \ra \Xi^{0}_c \g$  and $\Xi^{-'}_b \ra \Xi^{-}_b \g $
can  be predicted through a finite loop calculation. 
From
\beq
\G (S\ra T \g )=16 \a_{em}\,\fr{ E_\g^3 M_T}{  M_{S}} \, |A_{M1}|^2 \ ,
\eeq
we find
\bea
\Gamma(\Xi^{0'}_c)&=& (1.2\pm 0.7) \; {\rm KeV} \ , \nn \\
\Gamma(\Xi^{-'}_b)&=& (3.1\pm 1.8) \; {\rm KeV} \ .
\label{eq:xi}
\eea
%

Again the dominant  error in Eq.~(\ref{eq:xi}) is given by the uncertainty of 
$g_{2,3}$.

As in section~\ref{sec:S} in
 order to estimate the importance of chiral corrections we   use the ratios
(see Eq.~(\ref{eq:amt}) and~(\ref{eq:ams}))
\beq  
R(B^{(*)})= 
\fr{g_2 g_3 \Delta_{ST}a_g(B^{(*)})}{16 \pi f_\pi a_\chi(B^{(*)}) 
|c_{ST}(\mu)|} \ .
\eeq
We find (we  consider $ \La_\chi/2<\mu<\La_\chi$)
\bea
\displaystyle
R(\S_{c}^{+(*)})=R(\S_{b}^{0(*)})&=& - {(1.6 \pm 0.6)}/{|c_{ST}(\mu)|}, \nn \\
\displaystyle
 R(\Xi_{c}^{+'(*)})=R(\Xi_{b}^{0'(*)}&=&- {(2.4 \pm 0.8)}/{ |c_{ST}(\mu)|}
\eea
Therefore the
one loop chiral contribution   cannot be neglected for 
$|c_{ST}(\mu)|\sim {\cal O} (1)$.
In refs.~\cite{ivan} and \cite{chengnu,gnec},
numerical values for  all $S^{(*)}_c\ra T\g$ at ${\cal O}(1/\La_\chi)$
 are given using respectively the relativistic three quark model and the constituent quark-model.
As in section~\ref{sec:S}  we can define
\beq
c_{ST}(\mu)_{\overline{MS}}=c_{ST}^0+
\displaystyle{\fr{\left(c_{ST}^1(\mu)\right)_{\overline{MS}}}{\La_\chi}}
\ .
\eeq
From ref.~\cite{chengnu,ivan,gnec} we find 
\beq
0.83<|c_{ST}^0|<1.6 \ .
\eeq

Our results in Eq.~(\ref{eq:xi})  can be  compared
with other estimates existing  in the literature.
Ref.~\cite{ivan} reports 
$\Gamma(\Xi^{0'}_c)=0.17\pm 0.02 \ {\rm KeV} $,
while ref.~\cite{chengnu} quotes
$\Gamma(\Xi^{0'}_c)=0.3 \ {\rm KeV} $.
Further,  the  same argument as in section~\ref{sec:T} can be used 
also now to understand
 these  low values obtained  in  the
 constituent quark model~\cite{chengnu}.

For these decays the E2 amplitude is further suppressed than in the previous cases. 
The lowest--order contribution appears at 
${\cal O} (1/m_Q^3 \Lambda_{\chi}^2)$ and, therefore, can be neglected.

\section{Conclusions}
\label{sec:fin}

We have  calculated the electromagnetic  one photon decays 
$S^{*}\ra S \g $ and $S^{(*)} \ra T \g$
using  Heavy Hadron Chiral Perturbation Theory.
For each of these decays we have provided an  estimate of both the M1 and E2
amplitudes.
The computation of the M1 amplitudes up to  
${\cal O}(1/\Lambda_\chi^2)$ involves the introduction of   the unknown 
constants $c_S$ for $S^{*}\ra S \g $ and $c_{ST}$ for $S^{(*)} \ra T \g$.
Eliminating these couplings
we derive relations among different amplitudes.  
Moreover, since charm and bottom baryons are described by the same 
arbitrary constants, we can connect the amplitudes of the two kinds of hadrons.

The E2 contributions appear at  different higher orders  
for the three kinds of decays:
${\cal O}(1/\Lambda_\chi^2)$ for $S^{*}\ra S \g $,
${\cal O}(1/m_Q \Lambda_\chi^2)$ for $S^{*} \ra T \g$
and ${\cal O}(1/m^3_Q \Lambda_\chi^2)$ for $S \ra T \g$.
They introduce additional unknown constants: 
$c^{E2}_{S}$  for $S^{*}\ra S \g $; 
$c^{E2}_{T}$ and $ g'$ for $S^{*} \ra T \g$ 
(the E2 amplitude  for  $S \ra T \g$ is completely negligible).
The E2  effects can be strongly enhanced by  a term which is
infrared divergent in the exact chiral limit.
The possibility of measuring the ratio $A_{E2}/A_{M1}$,
using  polarized initial baryons, has been
suggested in ref.~\cite{savanu} and could  be performed with an 
analysis of the  photon distribution.

Furthermore, we obtain an absolute prediction for
$\G(\Xi^{0'(*)}_c\ra \Xi^{0}_c \g) $ and 
\mbox{$\G(\Xi^{-'(*)}_b\ra \Xi^{-}_b \g) $}.
At ${\cal O}(1/\Lambda_\chi^2)$, these decay widths 
do not get any contribution from local terms in the Lagrangian
and, therefore, their values are fixed by
a finite chiral loop calculation.

Finally, we have shown that chiral loops involving 
 photon--meson coupling
cannot be neglected in the computation of the amplitudes of these decays.
These interactions generate the dominant  contribution
to the electromagnetic decays of heavy baryons.

\section*{Acknowledgements}

I.S. wants to thank A. Della Riccia Foundation (Florence, Italy) for support. 
M.C.B. is indebted to the Spanish Ministry of Education and Culture 
for her fellowship.
This work has been supported in part by the European Union TMR Network  
``EURODAPHNE'' (Contract No. ERBFMX-CT98-0169), by DGESIC, Spain
(Grant No. PB97-1261) and by CICYT (Grant No. AEN99-0692).
We thank V.E. Lyubovitskij for some communications 
 concerning ref.~\cite{ivan}.

\end{document}